# (In,Ga)As gated-vertical quantum dot with an $Al_2O_3$ insulator

T. Kita[*] and D. Chiba

Semiconductor Spintronics Project, Exploratory Research for Advanced Technology, Japan Science and Technology Agency, 1-18 Kitamemachi, Aoba-ku, Sendai 980-0023, Japan and Laboratory for Nanoelectronics and Spintronics, Research Institute of Electrical Communication, Tohoku University, 2-1-1 Katahira, Aoba-ku, Sendai 980-8577, Japan

[*]Electronic address: tkita@riec.tohoku.ac.jp

Y. Ohno

Laboratory for Nanoelectronics and Spintronics, Research Institute of Electrical Communication, Tohoku University, and CREST, Japan Science and Technology Agency, 2-1-1 Katahira, Aoba-ku, Sendai 980-8577, Japan

H. Ohno

Laboratory for Nanoelectronics and Spintronics, Research Institute of Electrical Communication, Tohoku University, 2-1-1 Katahira, Aoba-ku, Sendai 980-8577, Japan and Semiconductor Spintronics Project, Exploratory Research for Advanced Technology, Japan Science and Technology Agency, 1-18 Kitamemachi, Aoba-ku, Sendai 980-0023, Japan

**Corresponding author:** Hideo Ohno

Laboratory for Nanoelectronics and Spintronics, Research Institute of Electrical Communication, Tohoku University, Katahira 2-1-1, Aoba-ku, Sendai 980-8577, Japan

tel/fax: +81-22-217-5553

e-mail: ohno@riec.tohoku.ac.jp

Abstract

We fabricated a gated-vertical (In,Ga)As quantum dot with an $Al_2O_3$ gate insulator deposited using atomic layer deposition and investigated its electrical transport properties at low temperatures. The gate voltage dependence of the *dI/dV-V* characteristics shows clear Coulomb diamonds at 1.1K. The metal-insulator gate structure allowed us to control the number of electrons in the quantum dot from 0 to a large number estimated to be about 130.

Gated-vertical quantum dots (QD) using (In,Ga)As-based resonant tunneling diode (RTD) structures have provided ideal systems to investigate various electron-spin correlation phenomena such as the Kondo effect and the Hund's first rule [1-3]. In such a tunable QD device based on III-V semiconductors, the Schottky gate formed around the QD plays a crucial role to control the energy levels and the number of electrons in the QD. Although it is very interesting to extend this technique to material systems such as narrow gap semiconductors that have different physical properties like an enhanced effective g-factor, a small effective mass, and a greater spin-orbit coupling, the low Schottky barrier height often observed in these systems have precluded direct application of the Schottky gate technique. Even in GaAs-based structures, the range of the applied gate voltage becomes rather limited when the doping concentration of the lead layers is increased. Thus development of a gate-structure that allows one to apply high electric-fields regardless of the material and doping concentration is much needed to expand the horizon of the physics probed by the gated-quantum dots. Atomic layer deposition (ALD) is a thin-film deposition technique based on self-limiting surface reactions [4]. It enables the thickness control on an atomic scale, is consistent with the conventional lithography techniques, and does not damage the surface of the semiconductors. In addition, thin-insulator films grown by ALD have been shown to have a large and uniform breakdown electric field [5-6] even for narrow-gap semiconductors [7]. In this Letter, we report the fabrication of a vertical quantum dot with an $Al_2O_3$ insulator deposited by ALD and demonstrate modulation of the number of electrons in a vertical QD from 0 to approximately 130.

The double barrier RTD structure was grown on an n-GaAs (001) substrate using molecular beam epitaxy. The RTD consists of an undoped 12 nm-$In_{0.05}GaAs_{0.95}As$ quantum well (QW) and two undoped $Al_{0.22}Ga_{0.78}As$ barriers of 9.0 nm (top) and 7.5 nm (bottom) thickness, respectively. 200 nm-thick n-GaAs (Si doping density of $1.5\times10^{17}$ $cm^{-3}$) lead layers were formed above and below the RTD. In order to form a non-alloyed ohmic contact to the top of the RTD structure, a 50 nm n-GaAs ($2\times10^{18}$ $cm^{-3}$), a 50 nm n-$In_xGa_{1-x}As$ graded layer ($x = 0 \sim 0.2$, $2\times10^{18}$ $cm^{-3}$), and a 20 nm n-$In_{0.2}Ga_{0.8}As$ ($1\times10^{19}$ $cm^{-3}$) were successively grown. The two-dimensional electron density in the $In_{0.05}Ga_{0.95}As$ QW is determined as $1.0 \times 10^{11}$ $cm^{-2}$ from the Shubnikov-de Haas oscillation of the magnetotransport measurement using a 10 μm-diameter RTD control sample made from the same wafer [8].

Figure 1 shows an oblique secondary electron microscope image of our vertical QD device with an $Al_2O_3$ gate insulator. The mesa-structure, consisting of an 800 nm diameter quantum dot and a 150-nm-wide mesa-line, was shaped by reactive ion etching using a 100 nm Au/Cr metal mask defined by electron-beam lithography. The mesa-line plays a role of a lead electrode (labeled as "drain") connected to the upper part of the RTD. Parallel conduction in the thin mesa-line is suppressed by the full depletion of the lead layers [9, 10]. A schematic cross-sectional view of this device is shown in the inset of Fig. 1. The Ohmic electrode on the backside of the substrate is labeled as "source". After removal of a 50 nm surface layer by wet chemical etching with a $H_3PO_4:H_2O_2:H_2O$ solution, 20 nm-thick $Al_2O_3$ gate insulator was deposited at substrate temperature of 300°C using alternating 0.1 s pulses of $Al(CH_3)_3$ and $H_2O$ with $N_2$ purges between steps. After deposition, the surface was covered entirely with an $Al_2O_3$ insulator. Then, a 50 nm-thick Au/Cr gate electrode was obliquely deposited by vacuum evaporation to ensure that the sidewall of the QW is covered by the gate metal. The dimension of the gate electrode is $30\times100$ $μm^2$. Finally, a contact hole for the drain was made by removing

Al$_2$O$_3$ on the mesa-line, and contact pads were formed.

Figure 2(a) shows the gate-source voltage ($V_G$) dependence of the drain-source current-voltage (*I-V*) characteristics measured at 22 K by using a cryogenic probe station. $V_G$ is varied from 4 V to −8 V in 4 V step. The zero-bias conductance is finite in the positive $V_G$ region, whereas the dot pinches off (number of electrons in the dot *n* = 0) for $V_G < -4$ V because the electron channel in QD is squeezed by negative $V_G$. Figure 2(b) shows the $V_G$ dependence of the gate leakage current density. The leakage current density is less than 10$^{-7}$ A/cm$^2$ in the entire −12 V to 10 V range and the breakdown electric field strength is 5~6 MV/cm, a value that is similar to that of the standard bulk Al$_2$O$_3$ deposited by ALD [6]. That value reflects the large breakdown field and high uniformity, even though it is grown on a vertical type device with three-dimensional asperity. The typical range of $V_G$ in a Schottky gate device is from −2 V to 0.5 V, limited by the breakdown of the Schottky gate (negative side) or the forward bias current (positive side), where the zero-bias depletion layer is 100~200 nm. However, the present device with a 20nm-thick Al$_2$O$_3$ insulator allows us to apply a gate voltage as high as 10 times greater than the Schottky gate, while keeping the distance between gate electrode and active channel about the same. Neglecting the effect of interface states, which we address later, this indicates that the insulator gate has a potential to modulate considerably wider range of electron densities.

The gate voltage dependence of the d*I*/d*V-V* characteristics was measured at 1.1 K in a refrigerator with liquid $^3$He/$^4$He mixture. The gate voltage was swept from ±10 V to ∓10 V at the rate of 8.2 mV/min, at which a one-way scan took about 40 hours. As shown by a color-scale plot in Fig. 3(a), clear Coulomb diamonds are observed up to V$_G$<-2 V, confirming the formation of a quantum dot. As $V_G$ > -0.2 V, the conductance oscillation reflecting the occurrence of single-electron tunneling can be seen in a broad range of $V_G$, whilst each peak cannot be resolved clearly at higher $V_G$. The electrons were fully depleted in the quantum dot and pinched-off at $V_G = -6.4$ V as no Coulomb diamond was observed with more negative gate voltage. In the $V_G > -6.4$ V region, the quantum dot holds one or more electrons. From the *dI/dV-V* spectra, we evaluated the charging energy for $n = 1$ to 2 to be about 5 meV, and those for $n = 3 \sim 8$ to be 1~2 meV, respectively. These values are similar to those in the Schottky-gated (In,Ga)As/(Al,Ga)As vertical QD structures [11].

The exact number of electrons in the quantum dot is obtained by counting the number of Coulomb peaks from the zero-electron state. As $V_G$ is increased well above the pinch-off voltage, however, the Coulomb diamonds became considerably smaller because of the reduction of the Coulomb energy. In the following, we estimate the effective channel diameter to obtain the maximum electron number (i.e. the number of electron in the quantum dot at $V_G = 10$V), assuming that the total current flowing across the quantum dot is proportional to the area of the electron channel in the high positive bias region. Figures 3(b) and (c) show the *I–V* characteristics of the control sample (10 μm diameter) and the gated QD device at $V_G = 10$ V, respectively. Supposing that the electron density in the QD at $V_G = +10$ V is the same as that in the QW of the reference sample ($1 \times 10^{11}$ cm$^{-2}$) and neglecting the influence of the lateral confinement, the effective diameter of the QD and the maximum electron number in the QD are estimated from the current ratio as about 400nm and 130, respectively.

Finally, we describe the hysteresis we see in the device. In Fig. 4, we plot the Coulomb diamond patterns for up and down sweeps of the gate voltage. When the gate voltage was increased from $-10$ V to 10 V (Fig. 4(a)), the pinch-off voltage was $-6.4$ V; whereas when it was decreased from 10 V to −10 V (Fig. 4(b)), the pinch-off voltage was −2.96 V. Therefore, the *dI/dV*-$V_G$ characteristics in this device show injection-type hysteresis. This suggests the existence of the interface states at the $Al_2O_3$/GaAs interface that trap electrons [12]. These interface states give rise to drift of the gate voltage at which the Coulomb peak appears, that may result in the change of the sizes of the Coulomb diamonds (for example see n=2 in Fig. 4) with changing the sweep direction of $V_G$. It should be noted that in the few-electron limit the averaged change in $V_G$ for *n* = 1 and 2 Coulomb diamonds at zero-bias is about 0.85 V. This value is 4~8 times greater than that of the conventional Schottky-gated QDs, which is typically 0.1 ~ 0.2 V [2], whilst the gate capacitance decrease is roughly 20% by inserting a 20 nm-thick $Al_2O_3$. The reason for the discrepancy is not fully understood, but it may be related to the interface states which could modify the shape of the confinement potential inside the QD.

Certainly we want to eliminate these interface states in future. However, when the amount of drift is slow enough and known, we can still perform a number of experiments within such a time window. For example, here we have a potential energy drift of 0.3 μeV/min with a sweep rate of 1.3 V/min from +6 V to 0 V, which allows us to perform measurements to address a number of QD properties. We also note that by optimizing the processing steps, we expect to be able to dramatically reduce the number of interface states [6].

In conclusion, we fabricated a vertical (In,Ga)As quantum dot structure with an $Al_2O_3$ gate insulator deposited by ALD. It showed clear Coulomb diamonds at low temperature and the gate voltage was controlled in the range of −12 V ~ 10 V. We have shown that the number of electrons in the quantum dot can be controlled from 0 to approximately 130. This device structure is a powerful tool to investigate the electron-spin correlation phenomenon in a large quantum dot and the quantum confinement property in various material systems that are not accessible using the regular Schottky gate structure.

The authors wish to acknowledge S. Tarucha, Y. Nishi, T. Hatano, and T. Yamaguchi for valuable discussions. They also thank K. Ohtani for discussion and his guidance. This work was partly supported by the Japan Society for the Promotion of Science.

**Figure captions**

Fig. 1 The scanning electron microscope (SEM) oblique image of the vertical quantum dot device with an $Al_2O_3$ gate insulator. The inset is an illustration of the cross-sectional view of the device.

Fig. 2 (a) The gate voltage ($V_G$) dependence of the drain-source *I-V* characteristics for $V_G$ = 4 V to −8 V with 4 V step. (b) The gate bias dependence of the gate leakage current density. The data were taken at 22 K.

Fig.3 (a) Color-scale plot of the gate voltage dependence ($V_G$ sweep rate of 8.2 mV/min) of d*I*/d*V*-*V* characteristics measured at 1.1 K when the $V_G$ was swept from -10V to 10V. (b) The *I–V* characteristics of the 10-µm-diameter control sample. (c) The *I-V* characteristics of the 800-nm-diameter quantum dot at $V_G$ = 10V.

Fig. 4 The Coulomb diamonds depended upon the sweep direction of the gate. (a) $V_G$ was swept from −10V to 10V, and (b) from 10V to −10V, respectively.

Fig1. T. Kita et al.

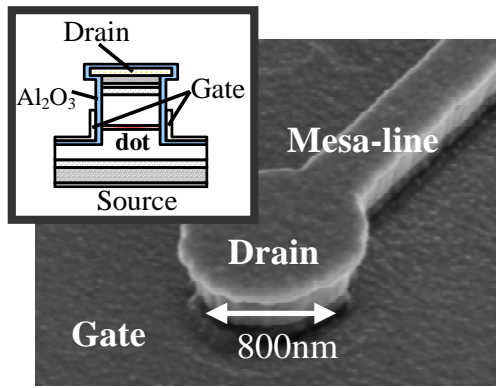

Fig2. T. Kita et al.

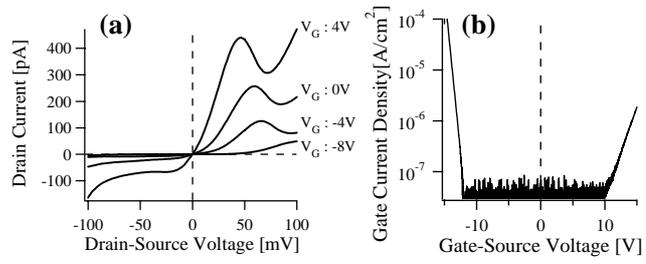

Fig3. T. Kita et al.

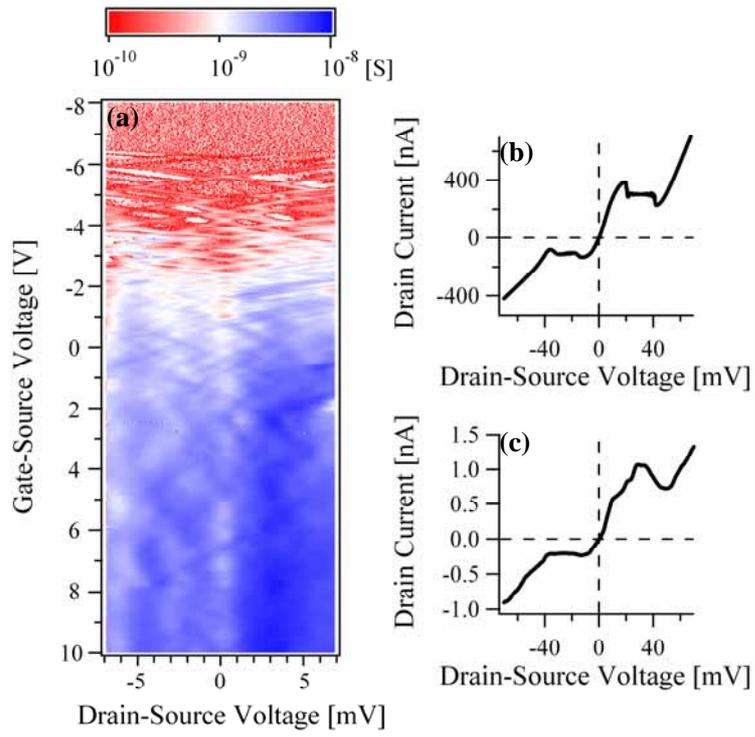

Fig4. T. Kita et al.

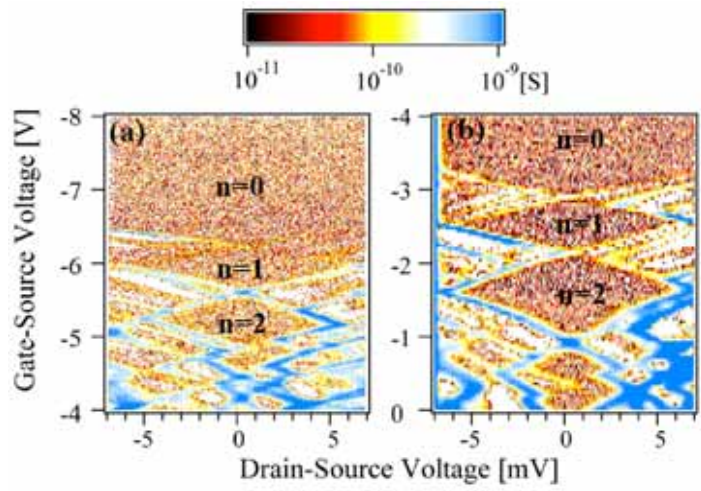